\documentstyle[11pt,aasms4]{article}

\newcommand{\postscript}[2]
 {\setlength{\epsfxsize}{#2\hsize}
  \centerline{\epsfbox{#1}}}

\begin{document}

\title{Correction for Blending Problem in Gravitational\\ 
       Microlensing Events by Using Hubble Space Telescope}
\author{Cheongho Han}
\affil{Dept.\ of Astronomy \& Space Science, \\
       Chungbuk National University, Cheongju, Korea 360-762}
\authoremail{cheongho@astro.chungbuk.ac.kr}


\begin{abstract}
The biggest uncertainty in determining microlensing 
parameters comes from the blending of source star images 
because the current 
experiments are being carried out toward very dense star fields:
the Galactic bulge and Magellanic Clouds.
The experiments try to correct the blending effects for 
individual events by introducing an additional lensing parameter, the 
residual flux, but this method suffers from very large uncertainties 
in the derived lensing parameters due to the degeneracies among 
the parameters.
In this paper, 
I propose to use the {\it Hubble Space Telescope} ({\it HST})
to correct blending effects.
With the high resolving power of the {\it HST} combined with the color
information from ground-based observations, one can uniquely identify
the lensed source star in the blended seeing disk, and thus 
the uncertainty in the derived time scale can be significantly reduced.
\end{abstract}

\vskip100mm

\bigskip
\bigskip
\centerline{submitted to {\it The Astrophysical Journal}: April 21, 1996}
\centerline{Preprint: CNU-A\&SS-01/97}

\newpage
\section{Introduction}

Presently, four groups are searching for Massive Astronomical 
Compact Halo Objects (MACHOs) by observing light variations of stars
caused by gravitational microlensing: MACHO (Alcock et al.\ 1995, 1997),
EROS (Aubourg et al.\ 1993; Ansari et al.\ 1997),
OGLE (Udalski et al.\ 1994a), and DUO (Alard 1996a).
From microlensing experiments toward Large Magellanic Cloud (LMC), 
one can constrain the MACHO dark-matter fraction in the Galactic halo. 
Experiments toward the Galactic bulge provide a new probe of 
low-mass stellar objects in the Galaxy.

To maximize the event rate, the searches are being carried out toward 
very dense star fields, where the images of stars are severely blended 
with one another, and thus observation is limited by crowding.
When an event in such a dense field is monitored, the observed light curve 
is affected by the light from stars that are not lensed:
blending effect.
If the effect is not properly corrected, the MACHO dark-matter fraction 
in the Galactic halo, determined from the lensing optical depth,
will be subject to great uncertainties.
In addition, blending makes identification of the major lens population 
difficult because it causes the determined time scale shorter 
than the true value.
The lensing optical depth represents the probability for 
a single source star to be lensed and is observationally determined by
$$
\tau = {\pi \over 2N_{\ast}T}
\sum_{i=1}^{N_{\rm event}} 
{ t_{{\rm E},i} \over \epsilon (t_{{\rm E},i})},
\eqno(1.1)
$$
where $\epsilon$ is the detection efficiency, 
$T$ is the total observation time, and 
$N_{\ast}$ and $N_{\rm event}$ are the total numbers
of monitored source stars and actually detected events, respectively.
The Einstein time scale is related to the physical parameters of a 
lens by
$$
t_{\rm E} = {r_{\rm E} \over v};\qquad
r_{\rm E} = \left(
{4GM_L \over c^2} {D_{\rm ol}D_{\rm ls} \over D_{\rm os}}
\right)^{1/2},
\eqno(1.2)
$$
where $r_{\rm E}$ is the Einstein ring radius, 
$v$ is the transverse speed of the lens relative to the observer-source
line of sight, $D_{\rm ol}$, $D_{\rm ls}$, and
$D_{\rm os}$ are the distances between the observer, lens, 
and the source, and $M_L$ is the mass of the lens.

When one of multiple stars in an {\it effective seeing disk}, the 
maximum undistinguishable separation between images, is gravitationally 
lensed, the measured flux is the sum of flux from the lensed star 
and residual flux from blended stars, 
$$
F = F_{0,i} A + B;\qquad B = \sum_{j\not = i} F_{0,j}.
\eqno(1.3)
$$
Here $F_{0,i}$ is the flux of the lensed star before or after 
gravitational amplification, and the 
amplification is related to lensing parameters by
$$
A(u) = {u^2 + 2 \over u (u^2+4)^{1/2} };\qquad 
u = \left[ \beta^2 + \left( {t-t_0 \over t_{\rm E} }\right)^2 
\right]^{1/2}.
\eqno(1.4)
$$
where $u$ is the lens-source separation in units of  $r_{\rm E}$,  
$\beta$ is the impact parameter, and $t_0$ is the time of maximum 
amplification.

There are two types of blending in microlensing.
The first case occurs when a star brighter than the detection limit, 
which is set by crowding, is lensed and the measured flux is 
affected by residual flux from other blended stars fainter than the 
detection limit.
The term `blending' in microlensing usually refers to this case, 
and I will refer to this type of blending as ``regular blending''.
The other type of blending occurs when one of several faint stars
below the crowding limit is lensed and its flux is associated with 
the flux from other stars in the effective seeing disk. 
In this case, the event appears seemingly with the brightest star 
as the source star because the lensed star is not resolved and 
thus not registered in the template plate. 
Due to the second type of blending, the relative probability of 
detecting low-amplification events (after proper efficiency correction) 
is higher than random distribution which is expected for unblended 
events.
Therefore, the second type of blending is often referred as 
``amplification bias''
(Blandford \& Narayan 1992; Narayan \& Wallington 1994; Nemiroff 1994;
Han 1997; Alard 1996b).

Blending, both regular blending and amplification bias, affects 
the results of the lensing experiments in various ways. 
First, blending restricts one to detecting only events with amplifications 
high enough to overcome the threshold residual flux level contributed by 
the blended stars in the effective seeing disk.  
For the detection of an event with a source star flux $F_{0,i}$, 
the amplification should satisfy the condition
$$
{
A_{\rm min} F_{0,i} + B \over 
F_{0,i} + B
} 
\geq 1.34,
\eqno(1.5)
$$ 
resulting in the minimum amplification for detection of
$$
A_{\rm min} = {1.34 (F_{0,i}+B) - B \over F_{0,i} }.
\eqno(1.6)
$$
Secondly, even for detected events one measures only the portion of 
the light curve which is above the threshold.
Since this part of the light curve mimics that of a shorter time scale 
event, what is measured is not the Einstein time scale $t_{\rm E}$,
but the effective time scale $t_{\rm eff}$.
The two time scales are related by
$$
t_{\rm eff} = \eta (\beta ) t_{\rm E};\qquad
\eta (\beta ) = \left[ 
\beta_{\rm max}^2 - \beta^2
\right]^{1/2}
\eqno(1.7)
$$
where $\beta_{\rm max} $ is the maximum allowed impact parameter
for detection and it is related to the minimum required 
amplification by
$$
\beta_{\rm max} = \left[
2\left( 1-A_{\rm min}^{-2} \right)^{-1/2} -2
\right]^{1/2}.
\eqno(1.8)
$$
Therefore, the observed time scale distribution is systematically 
shifted toward shorter time scales, i.e., $\eta \leq 1.0$.
In particular,  the time scales of amplification-biased events are 
much more seriously affected compared to those of regular-blended events
because of relaltively much higher residual fluxes. 
Since the lens mass scales as $M\propto t_{\rm E}^2$, 
the mass determined from the measured time scale will be underestimated 
and the lens population will be misinterpreted.
For example, blended events caused by main-sequence stars will be 
confused with unblended events caused by brown dwarfs.
The other important effect of amplification bias is that the 
optical depth to microlensing will be significantly 
overestimated  because events are generated by all stars, 
both resolved and unresolved, while $\tau$ is determined from only
the resolved stars (Han 1997). 
Therefore, if the blending is not properly corrected, 
the MACHO dark matter fraction of the Galactic halo will be 
seriously overestimated.

Although there have been several methods proposed to correct the 
blending problem, they are either applicable to limited cases,
suffers from large uncertainties in the determined lens parameters, or 
they are statistical not on individual event basis.
Currently, experiments try to correct the blending effects
by introducing an additional lensing parameter, i.e., the residual
flux $B$, into the light-curve fitting process  (Alcock et al.\ 1997).
However, this method suffers from increased uncertainties in derived
lensing parameters due to the degeneracies among them
and it sometimes results in physically unrealistic results, e.g.,
negative residual flux (M.\ R.\ Pratt 1996, private communication).    
A shift of star's image centeroid can be used to identify the lensed 
source star (Alard 1996a), but this method is applicable only
for a special event in which the template star is widely separated
from a lensed star. 
One can correct the blending effects statistically if the luminosity 
function of stars well below the detection limit can be constructed 
(Han 1997), but in this case we lose information about individual events.
Therefore, other methods should be explored.

In this paper, I propose to use the {\it Hubble Space Telescope} ({\it HST}) 
to correct blending effects in gravitational microlensing experiments.
With the high resolving power of the {\it HST} combined with color
information obtained from ground-based observation, one can 
uniquely identify the lensed source star in the effective seeing disk, 
and thus can significantly reduce the uncertainty in the derived 
time scale.

\section{Fraction of Faint Events}

The fields toward which the microlensing experiments are being carried
out are very crowded.
To estimate the averge number of stars blended in the 
effective seeing disk of a star, I construct a model luminosity 
function (LF) of the stars in the galactic bulge field toward which 
the blending is most severe.
I adopt the LF determined by J.\ Frogel (private communication) 
for stars brighter than the ground detection limit of $I_0=18.2$. 
For fainter stars in the range $18.2 \leq I_0 \leq 22.4$, 
I adopt the LF determined from {\it HST} observation by Light, Baum, 
\& Holtzman (1997).
For the part of LF even fainter than $I_0=22.4$, I adopt the LF of 
stars in the solar neighborhood which is determined by 
Gould, Bahcall, \& Flynn (1996).
The finally constructed LF is shown in Figure 1(a).
In the figure, the LF is normalized for stars in an effective seeing 
disk of size $1''\hskip-2pt .5$.\footnote{
The average seeing of the current experiment is $\sim 2''$.
However, if the angular separation between the lensed object and 
the template image is wide enough, the blending effect can 
be noticed by the shift of image centroid.
I, therefore, set the effective size of unseparable seeing disk 
to be $\theta_{\rm eff} = 1''\hskip-2pt .5$.
}
According to the model LF, there will be $\sim 36$ stars in average 
in this effective seeing disk.
Among them, the number of stars brighter than the detection limit 
comprises just $\sim 0.75$, and majority of them are 
stars fainter than the detection limit and thus cannot be resolved.

Among these stars in the effective seeing disk, the important source of 
blending comes from stars that are $\sim 2$ mag below the ground 
observation detection limit.
In the dense region of the galactic bulge, each resolved bright 
template star works effectively as multiple source stars 
due to amplification bias.
The effective number of source stars for a single template star 
with a flux $F_{\rm t}$ is determined by
$$
n_{\rm eff} = 1 + \int_0^{F_{\rm t}} dF \Phi (F) 
\beta_{\rm max} (F_{\rm t}, F) 
\left\langle {\epsilon (t_{\rm eff})/ \epsilon 
(t_{\rm E})}\right\rangle,
\eqno(2.1)
$$
where $\Phi (F)$ is the LF of stars in the field, and 
the factor $ \langle {\epsilon(t_{\rm eff})/ \epsilon (t_{\rm E})}\rangle $
is introduced to account for the decrease in the 
detection limit for shorter time scales.
Then the total number of effective source stars is obtained by 
integrating equation (2.1) over all stars brighter than the 
detection limit ($F_{\rm DL}$);
$$
N_{\rm eff} = 
\int_{F_{\rm DL}}^{\infty} dF_{\rm t} n_{\rm eff} \Phi (F_{\rm t})
$$
$$
=\int_{F_{\rm DL}}^{\infty} dF_{\rm t} \Phi (F_{\rm t}) + 
\int_{F_{\rm DL}}^{\infty} dF_{\rm t} \Phi (F_{\rm t})
\int_0^{F_{\rm t}} dF \Phi (F) 
\beta_{\rm max} (F_{\rm t}, F) 
\left\langle {\epsilon (t_{\rm eff})/ \epsilon (t_{\rm E})}\right\rangle .
\eqno(2.2)
$$
Finally, the distribution of effective source star brightness of 
microlensing events is obtained by differentiating $N_{\rm eff}$ by $F$;
$$
dN_{\rm eff} (F) = 
\Phi (F_{\rm t})dF_{\rm t} + 
\int_{F_{\rm DL}}^{\infty}dF_{\rm t} \Phi (F_{\rm t})
\left[ \Phi (F) \beta_{\rm max} (F_{\rm t}, F)
\langle\eta (t_{\rm eff})/\eta (t_{\rm E})\rangle
dF \right] .
\eqno(2.3)
$$
In equation (2.3), the first term represents the LF of template stars,
while the second term accounts for the additional source stars due to 
amplification bias.

For the computation of $ dN_{\rm eff} (F)$, I assume the detection 
efficiency is related to the time scale by a power law:
$ \epsilon \propto t_{\rm eff}^p$.
Since the lensing events caused by amplification bias have very short 
time scales ($\lesssim 10\ {\rm days}$) where the dependence of detection 
efficiency to time scale is close to linear, I adopt $p=1$.
With this approximation and the model LF, the brightness distribution 
of effective source stars is computed by equation (2.3) and shown 
in Figure 1(b).
One finds the contribution to the total events by the 
amplification-biased events is important; they comprise $\sim 40\%$ 
of the total events.
One arrives at a similar result for a different model of detection 
efficiency; for the case $p=1/2$,
$\sim 50\%$ of the total events are amplification biased [see Figure 1(c)]. 
One also finds that most of all these amplification-biased events 
comes from the part of the LF which is $\sim 2$ magnitudes below the 
detection limit, i.e., $18.2 \lesssim I_0 \lesssim 20$ 
(shaded regions in Figure 1).
According to the model LF there are $\sim 4$ such stars in each 
effective seeing disk.
There will be additional numerous very faint stars in the same effective 
seeing disk;  $\sim 30$ stars fainter than $I_0 \le 20$.
However, the contribution to the total event rate by these very faint 
stars will be negligible (see \S\ 4.1).

\section{Correction for Amplification-Biased Events}

In previous section, I showed that the contribution to the 
total microlensing events by faint blended source stars is substantial, 
and these sources mainly come from  $\sim 2$ magnitudes 
below the detection limit imposed by crowding limit.
In this section I show that most of these faint source stars can be 
resolved and precise measurements of their fluxes can be made
by using {\it HST} (see \S\ 3.1).
Once these stars are resolved in a single pair of {\it HST} images 
which are taken after events, one can isolate 
the lensed source star from other blended stars by using the color 
changes observed during the event from the ground.
This can be done by comparing the observed color curve to the 
one expected when each candidate source star is assumed to be
the lensed source (see \S\ 3.2).

\subsection{Resolving Faint Source Stars}

The {\it HST} delivers essentially diffraction-limited images (with a
seeing of $\theta_{\rm see} \sim 0''\hskip-2pt .1$ in $I$ band),
and thus it has more than $100$ times resolving power than the 
observations taken under favorable conditions from the ground
($\theta_{\rm see} \gtrsim 1''$).
Then source star detection is no longer limited by crowding but only
by photon statistics.
Therefore, the important sources of blending, stars with 
$\sim 2$ magnitudes below the detection limit, can be easily 
resolved and precise measurement of their unamplified fluxes 
can be made with a few minutes of post-event {\it HST} 
observation, i.e., with a photometric precision of $\lesssim 2\%$, 
corresponding to a signal-to-noise ratio of $S/N \gtrsim 50$ for 
$I = 24$ star.

Since the proposed {\it HST} observation is required only 
to resolve individual stars in the effective seeing disk and measure their 
brightnesses not to obtain lensing event light curve, 
it can be carried out after events.
However, the {\it HST} Wide Field Planetary Camera 2 (WFPC2) can cover 
only $3.25\ {\rm arcmin}^2$ of sky, while the area covered by the 
microlensing experiments is in scales of several ${\rm deg}^2$.
Therefore, observations are required for individual source stars.
In addition, multiple exposures will be occasionally required for some 
bright sources to prevent saturation of images.
The required observation time will be $\sim 10\ {\rm min}$
in each band, and thus it will take slightly less than an hour
including telescope operating time for each source star.
Since events are detected at the rate $\gtrsim 50\ {\rm yr}^{-1}$,
the required {\it HST} observation time will be 
$\sim 50\ {\rm hr}\ {\rm yr}^{-1}$.

\subsection{Isolating Lensed Source Star}

If the lensed source is identified and its flux is known, 
the precision of the time scale 
determination will be significantly improved.
To show the improvement in the precision of time scale 
determination with known source brightness, I carry out simulation 
of a representative event toward the galactic bulge where the 
blending is most severe.
In the simulation, the ground-based observation is assumed 
to use 1.27m telescope with a dichroic beam 
splitter to give simultaneous imaging in both $V$ and $R$ bands.
The CCD camera can detect $25\ {\rm photons}\ {\rm s}^{-1}$ with
a 1m telescope for $V=20$ star.             
This ground observation strategy is similar to that of MACHO group
(Alcock 1996).
There are four stars  in an effective seeing disk with an angular
radius $\theta_{\rm eff}/2;\ \theta_{\rm eff}=1''\hskip-2pt .5$.
The $V$- and $R$-band magnitudes of these stars are chosen 
based on the stellar distribution in the color-magnitude diagram 
provided by Alcock et al.\ (1997) and from the model LF, 
and they are listed in Table 1. 
In the table, each star is designated by $i=1$, 2, 3, and 4 according to 
their $V$-band brightnesses.
Among these stars, the truely lensed source is $i=2$ star, and the 
event has an observed effective time scale of $t_{\rm eff} = 16.5\ 
{\rm days}$.
I assume the flux of the blended image is measured with a moderate 
photometric precision of $f=5\%$ during the time span of 
$-3t_{\rm eff} \leq t \leq 3t_{\rm eff}$.
The image is monitored twice per night, and thus the total number of data
points is $6\times 16.5\times 2\sim 200$.

To correct the blending effect caused by the amplification bias, 
one should know which star among all stars in the effective seeing 
disk is lensed.
Theoretically, it is possible to determine the source star flux
from the shape of the light curve by introducing the residual
flux $B$ as an additional parameter in the light-curve
fitting process.
However, the increase in the number of parameters results in severe
degeneracy problem; with very different source star brightnesses 
different combinations of other lensing parameters can produce 
very similar light curves.
This degeneracy in lensing parameters is well illustrated in 
Figure 2(a) and 2(b).
In the figure, each light curve is 
is obtained by fitting the model light curve to the observed one
under the assumption that each star in the effective seeing disk is 
the lensed source star.
One finds that although the lensing parameters are very different 
with one another, the shapes of the light curves are very similar, resulting 
in very similar effective time scales of $t_{\rm eff} \sim 16.5$ days.
The best-fitting lensing parameters of individual light curves are 
listed in Table 1.

More practical method for the isolation of the lensed source star
is provided by measuring the color changes accompanied in blended events. 
In Figure 2(c), I show the expected {\it color curves}, color changes as 
a function of time, expected for individual candidate source stars.
For our example case where $i=2$ star is the lensed source, the 
color differences at the peak of the curves from those
of other curves range 0.05-0.15 magnitudes.
These differences are large enough to be easily measured from 
ground-based observation.                                     
Better photometry from the `Early Warning sytem' (Udalski et al.\ 1994b) 
and `Alert system' (Pratt et al.\ 1996) will make it possible to measure
even smaller color differences.

\subsection{Precise Time Scale Determination}

Then how much does the precision of the time scale determination 
improve with the proposed method compared to the conventional one.
For this comparison, I first compute the uncertainty of the time scale 
$t_{\rm E}$ determined by using the conventional method of correcting 
blending effect: four-parameter fitting. 
For this estimation, the observed light curves of the example events 
is fitted, including $B$, by using the relations (1.3) 
and (1.4), and the resulting values of $\chi^2$ as functions of 
$t_{\rm E}$ and $F_{0,i}$ are shown as a contour map in Figure 3. 
In the fitting process, I let the values of $\beta$ and $t_0$ vary so 
that the model curves under fixed values 
of $(t_{\rm E}, F_{0,i})$ fit best to the observed one.
The values of $\chi^2$ are computed by
$$
\chi^2 = \sum_{i=1}^{N_{\rm pt}} 
{
(N_{\rm T} -N_{\rm O})^2 
\over 
(fN_{\rm T})^2
},
\eqno(3.3.1)
$$
where 
$$
\cases{
N_{\rm pt}=200 & the number of data points \cr
N_{\rm T}(t) &   the photon counts of the model light curve \cr 
N_{\rm O} (t)&   the measured photon counts \cr
f=5\% &         the photometric precision of ground-based observation. \cr
}
$$
In Figure 3, the contours are drawn at the levels of
$\chi^2 = 1.0$, 4.0, and 9.0, i.e., $1\sigma$, $2\sigma$, and $3\sigma$ 
levels, from inside to outside, and 
the uncertainties of the recovered time scales for corresponding levels
are listed in Table 2.
One finds that when the source flux is not known, the uncertainty 
of the time scale is very large.
For our example event, the uncertainty of the obtained time scale is 
$28.1^{\rm d}\leq t_{\rm E} \leq 1412.5^{\rm d}$ 
measured at $1\sigma$ level while the true value is $t_{\rm E} = 
51.6^{\rm d}$.
On the other hand, if the lensed source star is identified from 
{\it HST} observation and its brightness is known,
the uncertainty in the recovered time scale decreases significantly.
For the example event, the uncertainty of the recovered time scale is
$47.7^{\rm d}\leq t_{\rm E} \leq 69.2^{\rm d}$ measured at the same level.

\section{Further Considerations}

\subsection{Very Faint Source Events}

For an event in which the lensed source is very faint,
it will be difficult to detect the source star in 
a short exposure post-event {\it HST} observation.
Even if these very faint stars are resolved with a deep {\it HST} image, 
identifying the lensed source star will be difficult due to 
mainly two reasons.
First, there are too many such faint stars in the effective seeing disk,
i.e., $\gtrsim 10$ stars in the range $20 \leq I_0 \leq 23$
according to the model LF.
Secondly, the stellar type of these stars is mostly main sequence stars 
with similar colors.
Therefore, it will be difficult to identify the lensed source star by 
using the color changes during the event.
In addition, even if the spectral types are quite different, 
the expected color changes from these faint star event will be small.

However, the contribution to the event rate by these very faint stars 
will be small.
This is because to be detected the star should be highly amplified, 
and thus rare.
When the efficiency model $\epsilon (t_{\rm eff})/\epsilon (t_{\rm E})= \eta$
is adopted,  very faint source events with source star brightnesses
$I_0 \ge 21$ comprise $\lesssim 5\%$ of the total events.

\subsection{Blending by Binary Stars}

Another complexity of blending arises when the source is composed of 
binary stars.
Source might exhibit the binary nature when both components are lensed,
but this case is very rare (Griest \& Hu 1992).
Typically, only one of the components are lensed and others works as stars 
contributing to the background flux: binary blending.

However, it is still possible to estimate the fraction of binary-blended 
events statistically.
Let $N_{\rm field}$ be the number of events blended by field (single) stars, 
and most of them will be identified from {\it HST} observations.
The remaining events that exhibit the effects of blending, but whose 
sources were not identified are most likely 
caused by binary blending.
Therefore, one can determine the binary-blended event fraction by
$$
f_{\rm bi} = {N_{\rm bi} \over N_{\rm tot}};
\qquad N_{\rm bi} = N_{\rm tot} - N_{\rm field}.
\eqno(4.2.1)
$$
Here $N_{\rm tot}$ is the total number of blended events 
by both binary companions and field stars, while $N_{\rm bi}$ is the number 
events blended by only binary companions. 
Therefore, {\it HST} observations of events that are blended by 
unrelated field stars can be bootstrapped to yield statistical estimation
of binary-blended events.
In addition, most binary systems are composed of one bright and 
the other relatively very faint star, causing minimal blending effects.

\acknowledgements
I acknowledge precious discussions with A.\ Gould.

\begin{table}
\begin{center}
\begin{tabular}{cccrlcc}
\hline
\hline
\multicolumn{3}{c}{lensed star} &
\multicolumn{3}{c}{lensing parameters} &
\multicolumn{1}{c}{ } \\
event&$V_i$(mag)&$R_i$ (mag)&$t_{\rm E}$ (days)&$\beta$&$t_0$ & $t_{\rm eff}$ (days) \\
\hline
$i=1$& 17.5   & 16.5    & 21.9\ \ \     & 0.3   & 0  & 16.2 \\
$i=2$& 19.0   & 18.3    & 51.6\ \ \     & 0.1   & 0  & 16.5 \\
$i=3$& 19.3   & 18.5    & 63.4\ \ \     & 0.08  & 0  & 16.5 \\
$i=4$& 20.0   & 19.2    & 113.3\ \ \    & 0.05  & 0  & 17.0 \\
\hline
\end{tabular}
\end{center}
\smallskip
\caption
{\footnotesize
The $V$- and $R$-band magnitudes of stars in the effective 
seeing disk for the example event.
In the effective seeing disk of the template star $i=1$, there are other 
three faint blended stars. 
Among these stars, the truely lensed source is $i=2$ star, and the
event has an observed time scale of $t_{\rm eff}=16.5$ days. 
Also listed are the lensing parameters which can produce very similar
light curves to the observed one under the assumption that each 
star is the lensed source.
}
\end{table}

\begin{table}
\begin{center}
\begin{tabular}{ccc}
\hline
\hline
\multicolumn{1}{c}{uncertainties} &
\multicolumn{1}{c}{with conventional} &
\multicolumn{1}{c}{with known source} \\
\multicolumn{1}{c}{level} &
\multicolumn{1}{c}{method} &
\multicolumn{1}{c}{star brightness} \\
\hline
$1\sigma$ & $1.45 \le \log t_{\rm E} \le 3.15$ & 
            $1.65 \le \log t_{\rm E} \le 1.84$ \\
\bigskip
          & $(28.1^{\rm d}\le t_{\rm E}\le 1412.5^{\rm d})$ 
	  & $(44.7^{\rm d}\le t_{\rm E}\le 69.2^{\rm d})$ \\
$2\sigma$ & $1.30 \le \log t_{\rm E} \le 3.30$ &
	    $1.56 \le \log t_{\rm E} \le 1.95$ \\
\bigskip
          & $(20.0^{\rm d}\le t_{\rm E}\le 1995.2^{\rm d})$ 
	  & $(36.3^{\rm d}\le t_{\rm E}\le 89.1^{\rm d})$ \\
$3\sigma$ & $1.13 \le \log t_{\rm E} \le 3.60$ &
	    $1.45 \le \log t_{\rm E} \le 2.05$ \\
          & $(13.5^{\rm d}\le t_{\rm E}\le 3981.1^{\rm d})$ 
	  & $(28.2^{\rm d}\le t_{\rm E}\le 112.2^{\rm d})$ \\
\hline
\end{tabular}
\end{center}
\smallskip
\caption
{\footnotesize
Uncertainties in the determined time scale of the example event.
The uncertainties in the second column are determined by the conventional 
method of four-parameter fitting, and they are compared to those, in the 
last column, determined with known source star brightness from 
{\it HST} observation.
}
\end{table}

\clearpage
\postscript{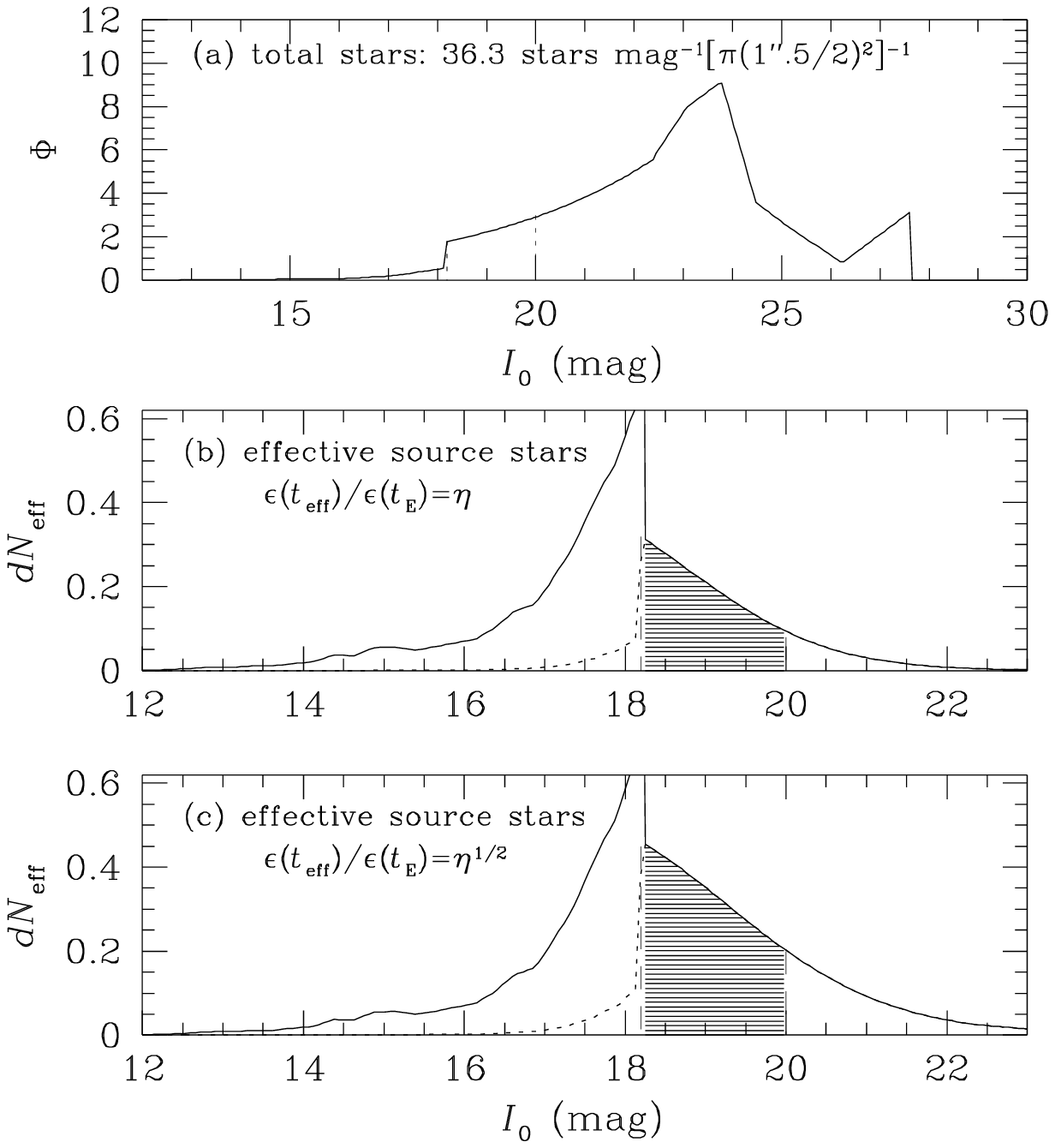}{0.75}
\noindent
{\small {\bf Figure 1:}\ 
(a): The model $I$-band luminosity function of stars in the galactic 
bulge field.   
There are totally $\sim 36\ {\rm stars}\ {\rm mag}^{-1}$ inside
the effective seeing disk of $\theta_{\rm eff} = 1''\hskip-2pt .5$. 
(b): The expected effective source star brightness distribution  
under the model that the detection efficiency is related to the 
time scale by
$\epsilon (t_{\rm eff})/\epsilon  (t_{\rm E}) = t_{\rm eff}/t_{\rm E}
=\eta$.
In this case the amplification-biased events comprises $\sim 40\%$
of the total events.
(c): The distribution of effective source star brightness
when $\epsilon (t_{\rm eff})/\epsilon  (t_{\rm E}) =\eta^{1/2}$.
Under this model the fraction of the amplification-biased events
to the total events is $\sim 50\%$.
}

\clearpage
\postscript{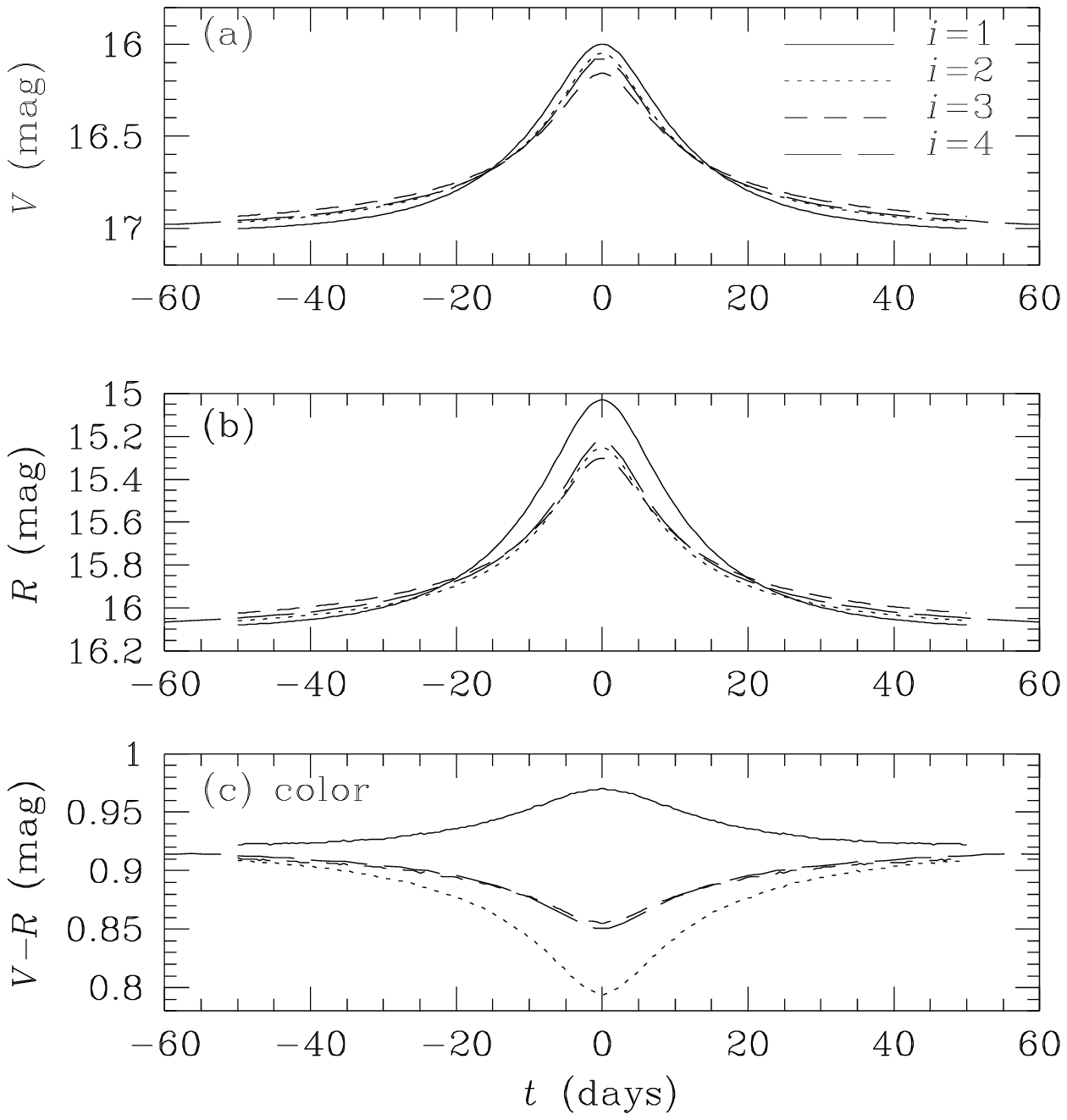}{0.75}
\noindent
{\small {\bf Figure 2:}\ 
The best fitting $V$- and $R$-band, (a) and (b) respectively, 
light curves to the observed one when individual stars 
in the effective seeing disk in Table 1 are assumed to be the 
lensed source.
Regardless of the source star brightnesses, all light curves 
results in similar effective time scale of $t_{\rm eff} = 16.5$ days.
Also shown, in panel (c), are the corresponding color curves for 
individual candidate source star events.
}

\newpage
\postscript{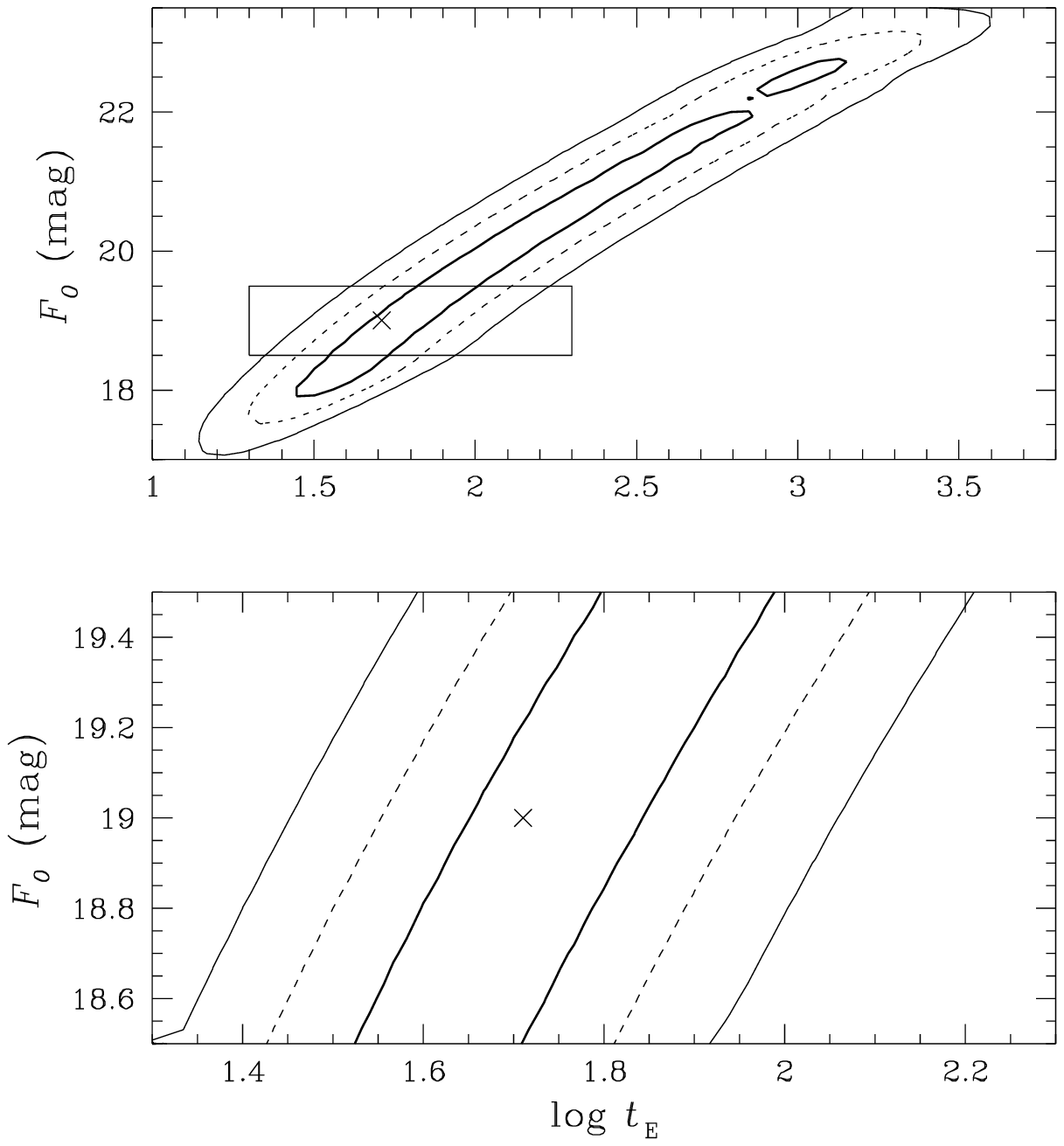}{0.75}
\noindent
{\small {\bf Figure 3:}\ 
The uncertainties of the best-fitting source star brightness 
and time scale for the example event.
The fit for a given value of $t_{\rm E}$ and $F_0$ is measured by
$\chi^2 $ and its value is shown as a contour map.
The contours are drawn at the levels of 
$\chi^2 =1.0$, 4.0, and 9.0 from inside to outside.
The region of parameter space in a small box in the upper 
panel is expanded and shown in the lower panel.
One finds that with known source flux the uncertainty 
of the derived time scale significantly reduces.
The ``$\times$'' mark represents the position of true source star 
brightness and time scale.
}

\end{document}